# Co-existence Between a Radar System and a Massive MIMO Wireless Cellular System

Stefano Buzzi, Marco Lops, Carmen D'Andrea, and Ciro D'Elia
University of Cassino and Southern Latium
Dept. of Electrical and Information Engineering, I-03043 Cassino (FR) – Italy
E-mail: {buzzi, lops, carmen.dandrea, delia}@unicas.it

*Abstract*—In this paper we consider the uplink of a massive MIMO communication system using 5G New Radio-compliant multiple access, which is to co-exist with a radar system using the same frequency band. We propose a system model taking into account the reverberation (clutter) produced by the radar system at the massive MIMO receiver. Then, we propose several linear receivers for uplink data-detection, ranging by the simple channel-matched beamformer to the zero-forcing and linear minimum mean square error receivers for clutter disturbance rejection. Our results show that the clutter may have a strong effect on the performance of the cellular communication system, but the use of large-scale antenna arrays at the base station is key to provide increased robustness against it, at least as far as data-detection is concerned.

## I. INTRODUCTION

The quest for ever increasing transmission rates in terrestrial communications has been pushing up the carrier frequencies towards bands traditionally occupied by radar: in particular, the $2-8$GHz bandwidth will be more and more overcrowded, whereby the issue of spectrum sharing between radar and communication systems has become a primary field of investigation [1]. Early studies on such a co-existence took a rather "radar-centric" approach - see, e.g., [2]–[4] - wherein the primary concern was to safeguard the received Signal-to-Interference plus Noise Ratio (SINR) at the radar receiver while limiting the amount of interference produced by the radar transmitter on a number of - possibly unlicensed - users. The focus has been recently steered back to the performance of the communication system as well, by introducing such techniques as co-design [5] and beamforming [6], while in [7] the reverberation produced by the radar transmitter onto the communication receiver (i.e., clutter or reflection produced by transient targets) has been recognized as a primary source of concern.

In the above context, the aim of the present contribution is to study the feasibility of having a search-radar system to co-exist with a fifth-generation (5G) wireless network, employing a standard Orthogonal Frequency Division Multiplexing (OFDM) modulation format and endowed with a *massive* MIMO array at the base station. Massive MIMO was introduced by Marzetta, in his pioneering paper [8]; this technology is currently arousing great interest in the scientific community [9], [10] and it will be widely employed in future cellular systems. Massive MIMO consists in using a very large number of service antennas (e.g., hundreds or thousands) in order to serve a lower number of mobile users with the time-division-duplex (TDD) protocol so as to exploit uplink/downlink channel reciprocity. In particular, our focus is on the effect that the massive structure may - or may not - have on clutter mitigation in the two relevant phases of the uplink haul, i.e. the *training phase* for user channel acquisition and the *demodulation phase* for data transmission. Following the 5G standard, we consider a Single-Carrier (SC) FDMA operating at a carrier frequency of 3GHz, and a co-existing radar system employing a *sophisticated* waveform with the same bandwidth. We present a model for the signal received at the Base Station (BS) array, accounting for the effect of the radar reflections on the whole set of packets entering the radar Pulse Repetition Time (PRT). We then propose several linear receivers for uplink data detection at the BS. Preliminary results show that using detection strategies at the BS based on the knowledge of the delays and directions of arrival of the clutter echoes, we can null the disturbance of the radar system on the cellular system though a simple linear processing, whereas the simple channel-matched beamforming is quite sensitive to the clutter power. Furthermore, the use of massive MIMO array at the BS, allows us to obtain an increase of system performance in terms of SINR, both in the case of knowledge of clutter statistic and in the case of channel-matched beamforming. Unfortunately, there is a notable gap between the attainable performance when the channel is assumed known and that observed when the channel is estimated though simple pilot-matched estimation, which suggests that more sophisticated channel estimation schemes are to be conceived.

This paper is organized as follows. In the next section we illustrate the considered system model, along with the model of the received signal at the BS, both in the case of uplink data transmission and uplink pilot transmission for channel estimation. Section III is devoted to the derivation of the considered linear uplink detection structures; Section IV contains numerical results, while, finally, concluding remarks are given in Section V.

## II. SYSTEM MODEL

Consider a single-cell massive MIMO communication system using SC-FDMA multiple access in the uplink, operating at a carrier frequency $f_c = 3$ GHz, and coexisting with a radar system using the same frequency band, as depicted in Fig. 1. With regard to the massive MIMO system, we use the following notation and assumptions:

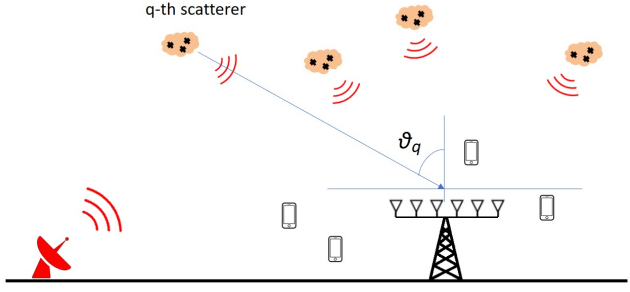

Fig. 1. A massive MIMO cellular system co-existing with a radar system. The BS received signal is corrupted by the clutter echoes. Ambient scatterers are seen as point-like targets placed at some random angles.

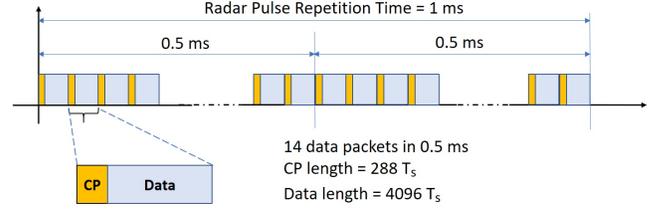

Fig. 2. Uplink frame structure. Data packets are made of a CP (of length 288 in discrete samples) and of information symbols (of length $N = 4096$ in discrete samples). The symbol time is such that 14 data packets fit into 0.5 ms. The radar PRT is 1 ms.

- $N$ denotes the number of subcarriers of the SC-FDMA system ($N = 4096$ will be assumed);
- The BS is equipped with a uniform linear array (ULA) with $M$ elements; fully digital beamforming is assumed, so that the number of RF chains coincides with the number of antennas.
- The mobile stations (MSs) transceivers are equipped with a single antenna, and the number of MSs in the system is $K$.
- The subcarrier spacing is denoted by $\Delta f$ ($\Delta f = 30$ KHz is assumed).
- A block fading channel is assumed with channel coherence bandwidth equal to $C\Delta f$, with $C = 16$. Otherwise stated, the channel can be considered constant over $C$ consecutive carriers and then takes a new value statistically independent from the previous one. Note that, for each user, and for each BS receive antenna, the channel state information amounts to $Q = N/C = 256$ complex scalar coefficients.
- The uplink channel between the $k$-th single-antenna MS and the BS on the $n$-th carrier is represented by an $M$-dimensional vector expressed as $\mathbf{h}_k^{(\lceil n/C \rceil)} = \beta_k \mathbf{g}_k^{(\lceil n/C \rceil)}$, where $\beta_k$ takes into account the path-loss and the log-normal shadowing (fully correlated across antennas and subcarriers), while $\mathbf{g}_k$ denotes the small-scale fading and is a random vector with $\mathcal{CN}(0, \mathbf{I}_M)$ distribution, with $\mathbf{I}_M$ the identity matrix of order $M$.
- The MSs transmit simultaneously using all the available subcarriers; user separation is performed in the spatial domain thanks to the use of a large number of antennas.
- The uplink frame structure is depicted in Fig. 2. Each packet is made of a cyclic-prefix (CP) and of a sequence of data symbols; the CP discrete length is $N_{\rm CP} = 288$, while the length of the data symbols is $N$. The timing is such that 14 packets fit into a 0.5 ms timeslot, which leads to a symbol time $T_s = 8.146$ ns. These numbers are inspired by the December 2017 3GPP first realease of the 5G New Radio standard.

With regard to the radar system, the following assumptions are made.

- The radar operates at the same carrier frequency as the wireless cellular system and it is assumed that there is full overlap between the bandwidths of the radar signal and of the communication signals transmitted by the MSs[1].
- The radar transmits a coded waveform, of duration $LT_s$; its baseband equivalent is expressed as

$$s_R(t) = \sqrt{P_T} \sum_{\ell=0}^{L-1} c_\ell \psi(t - \ell T_s), \quad (1)$$

wherein $P_T$ is the radar transmitted power, $[c_0, c_1, \ldots, c_{L-1}]$ is the unit-energy radar code, and $\psi(\cdot)$ is the base pulse; we assume that $\psi(\cdot)$ is a unit-energy rectangular pulse of duration $T_s$. The value $L = 32$ is assumed in this paper.
- The waveform $s_R(t)$ is transmitted periodically every $T_{\rm PRT} = 1$ ms, with $T_{\rm PRT}$ the PRT.

In the following, we provide a model for the uplink signal received at the BS, taking into account both the data signals transmitted by the MSs and the contribution from the radar system due to the presence of scatterers in the surrounding environment.

A block scheme of the generic MS transmitter is reported in the upper part of Fig. 3, while the lower part of the same figure represents a block scheme of the uplink receiver at the generic receive BS antenna. As it is seen from the frame structure in Fig. 2, 14 data packets fit into a 0.5 ms time window; some of these packets can be used to transmit known training symbols in order to enable channel estimation. In the following, we describe separately the signal model for the data packets in the training phase and in the data communication phase. We consider the latter situation first.

*A. Signal model during uplink data transmission*

Consider the generic $\ell$-th data packet; denote by $\mathbf{x}_k(\ell)$ an $N$-dimensional vector containing the data symbols from the $k$-th MS to be transmitted in the $\ell$-th data packet; denote by $\mathbf{X}_k(\ell)$ the $N$-dimensional vector representing the isometric FFT of $\mathbf{x}_k(\ell)$. Referring to the lower part of Fig. 3, it is easily shown that the observable corresponding to the $n$-th

---

[1]This assumption is made to simplify the notation; the generalization of the results of this paper case of partial overlap between the bandwidths can be treated with ordinary efforts.

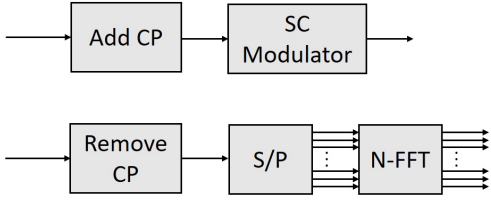

Fig. 3. Upper figure: Block-scheme of the transmitter at the generic mobile station. Lower figure: Block-scheme of the BS receiver at the generic antenna; assuming fully-digital beamforming at the BS, this scheme is to be replicated for each receive antenna.

subcarrier after the FFT operation can be represented through the following $M$-dimensional vector:

$$\mathbf{y}(\ell)^{(n)} = \sum_{k=1}^{K} \sqrt{p_k} \mathbf{X}_k(\ell)^{(n)} \mathbf{h}_k^{(\lceil n/C \rceil)} + \mathbf{W}(\ell)^{(n)} + \mathbf{C}(\ell)^{(n)}, \quad (2)$$

for $n = 1, \ldots, N$. In the above equation, $p_k$ is the power transmitted by the $k$-th MS, $\mathbf{X}_k(\ell)^{(n)}$ is the $n$-th entry of the vector $\mathbf{X}_k(\ell)$, $\mathbf{W}(\ell)^{(n)}$ is a $\mathcal{CN}(\mathbf{0}, \sigma_w^2 \mathbf{I}_M)$ random vector representing the additive thermal noise, while $\mathbf{C}(\ell)^{(n)}$ is the clutter contribution generated by the radar system on the $n$-th subcarrier; an expression for such vector will be given in the following. Grouping together the observable corresponding to the $N$ subcarriers we finally get the following $(M \times N)$-dimensional matrix for the observables corresponding to the $\ell$-th data packet:

$$\mathbf{Y}(\ell) = \sum_{k=1}^{K} \sqrt{p_k} \left( \left[ \mathbf{h}_k^{(1)} \ldots \mathbf{h}_k^{(Q)} \right] \otimes \mathbf{1}_{1 \times C} \right) \text{diag}(\mathbf{X}_k(\ell)) \\ + \mathbf{W}(\ell) + \mathbf{C}(\ell), \quad (3)$$

where $\otimes$ denotes kronecker product and $\mathbf{1}_{1 \times C}$ denotes a $C$-dimensional row vector with unit entries.

*B. Signal model during uplink training*

Consider now the case in which the MSs transmit known pilot sequences to enable channel estimation at the BS. Let $T$ denote the number of consecutive packets devoted to training, and let $\mathbf{p}_k(1), \ldots, \mathbf{p}_k(T)$ denote $N$-dimensional vectors containing the $k$-th MS pilots to be used in the $T$ packets used for channel estimation. Focusing on the $\ell$-th packet (with now $\ell = 1, \ldots, T$), and following the same steps as in the previous section, it is easily shown that the observable at the output of the FFT block at the BS receiver can be written as the following $(M \times N)$-dimensional matrix

$$\mathbf{Y}(\ell) = \sum_{k=1}^{K} \sqrt{p_k} \left( \left[ \mathbf{h}_k^{(1)} \ldots \mathbf{h}_k^{(Q)} \right] \otimes \mathbf{1}_{1 \times C} \right) \\ \text{diag}(\mathbf{W}_{N,FFT} \mathbf{p}_k(\ell)) + \mathbf{W}(\ell) + \mathbf{C}(\ell), \quad (4)$$

where, now, $\mathbf{W}_{N,FFT}$ is the $(N \times N)$-dimensional matrix performing an isometric FFT[2]. Assume now that the $M$-dimensional channel vectors $\mathbf{h}_k^{(q)}$, $\forall k = 0, \ldots, K-1$, are to be estimated; to this end, the columns from the $[(q-1)C+1]$-th to the $[qC]$-th of the matrices $\mathbf{Y}(1), \ldots, \mathbf{Y}(T)$ are to be picked; they form the following observable:

$$\mathcal{Y}_q = \sum_{k=1}^{K} \sqrt{p_k} \mathbf{h}_k^{(q)} \mathbf{P}_k^{(q)\,T} + \mathcal{W}_q + \mathcal{C}_q, \quad (5)$$

where

$$\mathcal{W}_q = \left[ \mathbf{W}(1)_{:,(q-1)C+1:qC} \cdots \mathbf{W}(T)_{:,(q-1)C+1:qC} \right],$$
$$\mathcal{C}_q = \left[ \mathbf{C}(1)_{:,(q-1)C+1:qC} \cdots \mathbf{C}(T)_{:,(q-1)C+1:qC} \right],$$

and $\mathbf{P_k}^{(q)}$ is a $(TC)$-dimensional vector defined as follows:

$$\mathbf{P}_k^{(q)} \triangleq \left[ (\mathbf{W}_{N,FFT}\mathbf{p}_k(1))_{(q-1)C+1:qC}, \\ \ldots, (\mathbf{W}_{N,FFT}\mathbf{p}_k(T))_{(q-1)C+1:qC} \right]. \quad (6)$$

[2] The $(m,n)$-th entry of $\mathbf{W}_{N,FFT}$ is thus $\frac{1}{\sqrt{N}} e^{-j2\pi(m-1)(n-1)/N}$.

*C. Clutter modeling*

We now illustrate the clutter model and provide an explicit expression for the $(M \times N)$-dimensional clutter matrix $\mathbf{C}(\ell)$ affecting the $\ell$-th received data packet.

The clutter disturbance is actually generated by a large set of discrete scatterers in the surrounding environment. Given the BS array dimension it is reasonable to assume that these scatterers are seen by the BS as "colocated" and, thus, the radar-to-BS channel can be modeled as a LTI channel with the following vector-valued impulse response:

$$\mathbf{h}(t) = \sum_{q=0}^{N_s-1} \sum_{m=0}^{Q-1} \beta_{q,m} \mathbf{b}(\theta_q) \delta(t - \tau_q - m/W). \quad (7)$$

In the above equation, $N_s$ denotes the number of scatterers in the surrounding environment that contribute to the clutter disturbance; $\theta_q$ is the direction of arrival of the clutter contribution from the $q$-th scatterer, $\mathbf{b}(\theta) = [1\ e^{-j2\pi d \sin(\theta)/\lambda} \ldots e^{-j(M-1)2\pi d \sin(\theta)/\lambda}]^T$ is the BS ULA array response; $\tau_q$ is the propagation delay associated to the signal generated from the $q$-th scatterer. Moreover, since the signal bandwidth $W$ exceeds the channel coherence time, we also assume that each physical scatterer generates $Q$ clutter echoes spaced apart by integer multiples of $1/W$; accordingly, $\beta_{q,m}$ is the reflection coefficient associated to the $m$-th replica from the $q$-th scatterer.

Now, recall that the radar transmits the waveform in (1); this waveform travels through a channel with impulse response $\mathbf{h}(t)$ and then is passed through a filter with a rectangular impulse response of duration $T_s$ and sampled at rate $1/T_s$. After A/D conversion, the baseband equivalent of the clutter disturbance can be represented as the following vector-valued discrete-time (at rate $1/T_s$) signal:

$$\widetilde{\mathbf{s}}_R(\eta) = \sum_{q=0}^{N_s-1} \sum_{m=0}^{Q-1} \sum_{p=0}^{L-1} \sqrt{P_T} \beta_{q,m} c_p \mathbf{b}(\theta_q) \\ r_\psi((\eta - p)T_s - m/W - \tau_q), \quad (8)$$

with $r_\psi(\cdot)$ the autocorrelation function of the base pulse.

Now, refer to the frame structure of Fig. 2 and assume, for simplicity, that the radar transmits its signal at the beginning of a 0.5 ms timeframe[3]. Denoting by $T_{\text{pkt}} = (4096 + 288)T_s$ the duration of a data packet including its CP, the generic $\ell$-th packet starts at time $\ell T_{\text{pkt}} + T_{\text{CP}}$ and ends at $(\ell+1)T_{\text{pkt}}$. Let now $\mathcal{S}(\ell)$ denote the set of the scatterers corrupting the reception of the $\ell$-th data packet. Since the contribution from the generic $q$-th scatterer starts at $\tau_q$ and stops at $\tau_q + QT_s + LT_s$, it is easily seen that the set $\mathcal{S}(\ell)$ can be defined as

$$\mathcal{S}(\ell) = \{q \in \{0, 1, \ldots, N_s - 1\} : \\ [\tau_q, \tau_q + QT_s + LT_s] \cap [\ell T_{\text{pkt}} + T_{\text{CP}}, (\ell+1)T_{\text{pkt}}] \neq \emptyset\}, \quad (9)$$

with $\emptyset$ denoting the empty set. Using the above notation, it can be easily shown that the clutter $(M \times N)$-dimensional matrix appearing in Eqs. (3) and (4) can be expressed as

$$\mathbf{C}(\ell) = \sum_{q \in \mathcal{S}(\ell)} \sum_{m=0}^{Q-1} \sum_{p=0}^{L-1} \sqrt{P_T} \beta_{q,m} c_p \mathbf{b}(\theta_q) \mathbf{r}_{q,p,m}^T(\ell) \mathbf{W}_{N,FFT}, \quad (10)$$

wherein

$$\mathbf{r}_{q,p,m}(\ell) = \left[ r_\psi \left( \ell T_{\text{pkt}} + T_{\text{CP}} + T_s - pT_s - \tfrac{m}{W} - \tau_q \right), \\ \ldots, r_\psi \left( (\ell+1)T_{\text{pkt}} - pT_s - \tfrac{m}{W} - \tau_q \right) \right]^T, \quad (11)$$

and

$$\widetilde{\mathbf{R}}_{q,\ell,m} = \sum_{p=0}^{L-1} c_p \mathbf{r}_{q,p,m}^T(\ell) \mathbf{W}_{N,FFT}.$$

## III. RECEIVER PROCESSING

In this section we focus on the signal processing algorithms at the BS to estimate the uplink channels and decode the MSs data symbols. In the following, we assume knowledge of the delays $\tau_q$ and directions of arrival $\theta_q$ of the clutter echoes. The design of adaptive procedures for automatic estimation of the clutter second-order statistics is a topic certainly worth being investigated but out of the scope of this paper.

### A. Uplink channel estimation

Given the data model (5), a simple estimator for the channel vector $\mathbf{h}_k^{(q)}$, $\forall k, q$, is obtained through the following pilot-matched (PM) processing

$$\widehat{\mathbf{h}}_k^{(q)} = \mathcal{Y}_q \frac{\mathbf{P}_k^{(q)}}{\sqrt{p_k} \left\| \mathbf{P}_k^{(q)} \right\|^2}. \quad (12)$$

Alternatively, more sophisticated channel estimation strategies can be proposed, which we omit here due to lack of space.

### B. Uplink data detection

Consider now the problem of data detection. For the sake of simplicity, we process separately the columns of the received matrix $\mathbf{Y}(\ell)$; recall that its $n$-th column is expressed as in (2), wherein it is easily shown that the covariance matrix of the clutter vector $\mathbf{C}(\ell)^{(n)}$ is expressed as

$$\mathbf{K}_{\mathbf{C}(\ell)^{(n)}} = \sum_{q \in \mathcal{S}(\ell)} \sum_{m=0}^{Q-1} P_T \sigma_\beta^2(q,m) \left| \widetilde{\mathbf{R}}_{q,\ell,m}^{(n)} \right|^2 \mathbf{b}(\theta_q) \mathbf{b}^H(\theta_q). \quad (13)$$

[3]This assumption can be removed with ordinary efforts.

Now, in order to detect the data symbols $\mathbf{X}_k(\ell)^{(n)}$ based on the data model (2), several detection strategies can be envisaged. For the sake of simplicity we focus here on linear detection rules and in particular we consider four possible strategies.

*1) Channel-matched beamforming (CM):* Based on the channel estimate $\widehat{\mathbf{h}}_k^{(\lceil n/C \rceil)}$, a soft estimate of $\mathbf{X}_k(\ell)^{(n)}$ is built as

$$\widehat{\mathbf{X}}_k(\ell)^{(n)} = \frac{\widehat{\mathbf{h}}_k^{(\lceil n/C \rceil) H} \mathbf{y}^{(n)}(\ell)}{\sqrt{p_k} \left\| \widehat{\mathbf{h}}_k^{(\lceil n/C \rceil)} \right\|^2}. \quad (14)$$

*2) Zero-Forced clutter (ZF):* This receiver zero-forces the clutter contribution by projecting the data vector along a direction that is orthogonal to the subspace spanned by the clutter covariance matrix $\mathbf{K}_{\mathbf{C}(\ell)^{(n)}}$. Letting $\mathbf{U}(\ell)^{(n)}$ be a matrix containing the eigenvectors of the matrix $\mathbf{K}_{\mathbf{C}(\ell)^{(n)}}$ associated to non-zero eigenvalues, we have in this case

$$\widehat{\mathbf{X}}_k(\ell)^{(n)} = \frac{\left[ \left( \mathbf{I}_M - \mathbf{U}(\ell)^{(n)} \mathbf{U}(\ell)^{(n) H} \right) \widehat{\mathbf{h}}_k^{(\lceil n/C \rceil)} \right]^H \mathbf{y}^{(n)}(\ell)}{\sqrt{p_k} \left\| \left( \mathbf{I}_M - \mathbf{U}(\ell)^{(n)} \mathbf{U}(\ell)^{(n) H} \right) \widehat{\mathbf{h}}_k^{(\lceil n/C \rceil)} \right\|^2}. \quad (15)$$

*3) Linear minimum mean square (LMMSE) data detector:* In this case, we have

$$\widehat{\mathbf{X}}_k(\ell)^{(n)} = \sqrt{p_k} \widehat{\mathbf{h}}_k^{(\lceil n/C \rceil) H} \mathbf{K}_{\mathbf{y}(\ell)^{(n)}}^{-1} \mathbf{y}^{(n)}(\ell), \quad (16)$$

with

$$\mathbf{K}_{\mathbf{y}(\ell)^{(n)}} = \sum_{j=1}^K p_j \widehat{\mathbf{h}}_j^{(\lceil n/C \rceil)} \widehat{\mathbf{h}}_j^{(\lceil n/C \rceil) H} + \sigma_w^2 \mathbf{I}_M + \mathbf{K}_{\mathbf{C}(\ell)^{(n)}}. \quad (17)$$

*4) Full zero-forcing (FZF):* This receiver zero-forces both the clutter contribution and the multi-user interference. In order to detect $\mathbf{X}_k(\ell)^{(n)}$, let $\mathbf{U}_k(\ell)^{(n)}$ be a matrix containing the eigenvectors of the matrix $\mathbf{K}_{\mathbf{y}(\ell)^{(n)}} - \sigma_w^2 \mathbf{I}_M - p_k \widehat{\mathbf{h}}_k^{(\lceil n/C \rceil)} \widehat{\mathbf{h}}_k^{(\lceil n/C \rceil) H}$ associated to non-zero eigenvalues. The data estimator is written as

$$\widehat{\mathbf{X}}_k(\ell)^{(n)} = \frac{\left[ \left( \mathbf{I}_M - \mathbf{U}_k(\ell)^{(n)} \mathbf{U}_k(\ell)^{(n) H} \right) \widehat{\mathbf{h}}_k^{(\lceil n/C \rceil)} \right]^H \mathbf{y}^{(n)}(\ell)}{\sqrt{p_k} \left\| \left( \mathbf{I}_M - \mathbf{U}_k(\ell)^{(n)} \mathbf{U}_k(\ell)^{(n) H} \right) \widehat{\mathbf{h}}_k^{(\lceil n/C \rceil)} \right\|^2}. \quad (18)$$

## IV. NUMERICAL RESULTS

In our simulation setup the system parameters detailed in Section II are used, while the channel vectors between the BS and the MSs are generated through the superposition of small-scale Rayleigh distributed fading (independent across antennas at the BS large array), log-normal shadowing, and distance-dependent path-loss – the three slope path loss model detailed in reference [11] is used. The MSs distance from the BS is uniform in the range $[20, 500]$ m., the additive thermal noise is assumed to have a power spectral density of -174 dBm/Hz, and the front-end receiver is assumed to have a noise figure of 3 dB.

We report results for both the case of perfect channel state information (CSI) and for the case in which the PM channel estimation detailed in Section III-A is used, with $T = 7$. The MSs transmit power is set at $100$ mW, both in the training and data transmission phases.

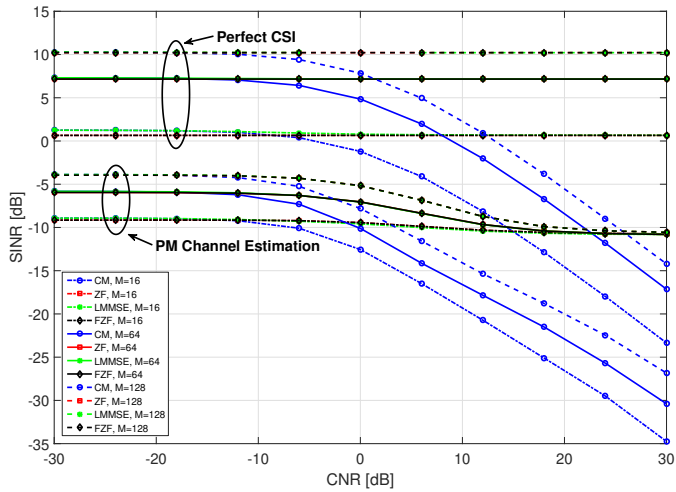

Fig. 4. SINR versus CNR of four detection strategies with $K = 1$ and different values of $M$.

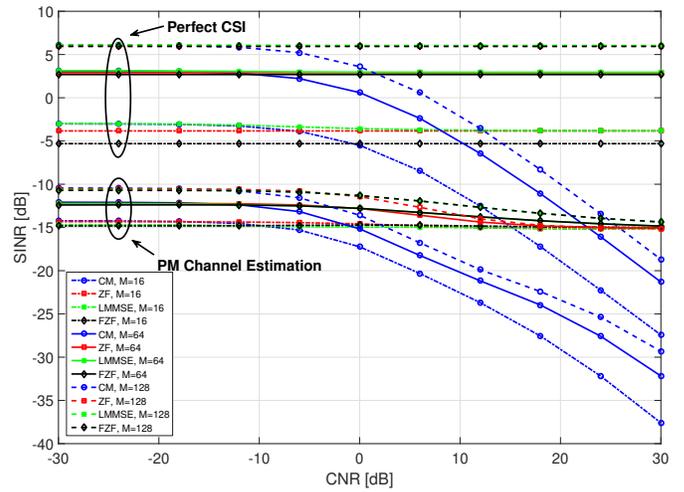

Fig. 5. SINR versus CNR of four detection strategies, with $K = 5$ and different values of $M$.

In Fig. 4 we report the SINR versus the Clutter-to-Noise Ratio (CNR) for four detection stategies in the single-user case, for different number of BS antennas $M$. Fig. 5 shows the same results as Fig. 4 for the multiuser case ($K = 5$ is assumed). Inspecting the figures, we observe the following: (i) the PM channel estimation procedure introduces a severe performance degradation, mainly due to the fact that channel estimation happens with a too short pilot sequence and with a too low SNR; this suggests that more sophisticated channel estimation schemes should be considered; (ii) of the considered data-detection structures, only the CM receiver is sensitive to an increase in the CNR, while the other receivers are actually robust to the clutter echoes; and (iii) for all the considered detection structures, it is seen that increasing the number of antennas at the BS provides a considerable gain and robustness, also with respect to the radar interference: increasing the antenna array size from $M = 16$ to $M = 128$ is shown in the figures to lead to a performance increase of 9.6 dB in the received SINR. Our results thus, although being preliminary, show that a cellular system may be made resilient to the interference produced by a radar system.

## V. Conclusions

The paper has considered a single-cell massive MIMO communication system using SC-FDMA multiple access in the uplink and coexisting with a radar system using the same frequency band. After the derivation of the system and signal model corresponding to this novel scenario, several receivers have been proposed and analyzed, based on the knowledge of the radar signal covariance matrix, showing that using a large number of antennas at the BS provides some robustness against the radar interfering signal. Further research is needed of course, in particular about the design of adaptive data detectors that do not require knowledge of the radar signal covariance matrix, as well as about the consideration of more sophisticated channel estimation schemes.

### Acknowledgement

The paper has been supported by the MIUR program "Dipartimenti di Eccellenza 2018-2022".

### References

[1] H. Griffiths, L. Cohen, S. Watts, E. Mokole, C. Baker, M. Wicks, and S. Blunt, "Radar spectrum engineering and management: technical and regulatory issues," *Proceedings of the IEEE*, vol. 103, no. 1, pp. 85–102, 2015.
[2] A. Aubry, A. De Maio, M. Piezzo, and A. Farina, "Radar waveform design in a spectrally crowded environment via nonconvex quadratic optimization," *IEEE Trans. Aerosp. Electron. Syst.*, vol. 50, no. 2, pp. 1138–1152, 2014.
[3] A. Aubry, A. De Maio, Y. Huang, M. Piezzo, and A. Farina, "A new radar waveform design algorithm with improved feasibility for spectral coexistence," *IEEE Transactions on Aerospace and Electronic Systems*, vol. 51, no. 2, pp. 1029–1038, 2015.
[4] H. Deng and B. Himed, "Interference Mitigation Processing for Spectrum-Sharing Between Radar and Wireless Communications Systems," *IEEE Trans. Aerosp. Electron. Syst.*, vol. 49, no. 49, pp. 1911–1919, 2013.
[5] B. Li and A. Petropulu, "Spectrum sharing between matrix completion based MIMO radars and a MIMO communication system," in *Acoustics, Speech and Signal Processing (ICASSP), 2015 IEEE International Conference o*. IEEE, 2015, pp. 2444–2448.
[6] J. A. Mahal, A. Khawar, A. Abdelhadi, and T. C. Clancy, "Spectral Coexistence of MIMO Radar and MIMO Cellular System," *IEEE Trans. Aerosp. Electron. Syst.*, vol. 53, no. 2, pp. 655–668, 2017.
[7] L. Zheng, M. Lops, X. Wang, and E. Grossi, "Joint Design of Overlaid Communication Systems and Pulsed Radars," *IEEE Trans. Signal Processing*, vol. 66, no. 1, pp. 139–154, 2018.
[8] T. L. Marzetta, "Noncooperative cellular wireless with unlimited numbers of base station antennas," *IEEE Transactions on Wireless Communications*, vol. 9, no. 11, pp. 3590–3600, Nov. 2010.
[9] E. Björnson, J. Hoydis, and L. Sanguinetti, "Massive MIMO networks: Spectral, energy, and hardware efficiency," *Foundations and Trends in Signal Processing*, vol. 11, no. 3-4, pp. 154–655, 2017. [Online]. Available: http://dx.doi.org/10.1561/2000000093
[10] T. L. Marzetta, E. G. Larsson, H. Yang, and H. Q. Ngo, *Fundamentals of Massive MIMO*. Cambridge University Press, 2016.
[11] S. Buzzi and C. D'Andrea, "Cell-free massive MIMO: User-centric approach," *IEEE Wireless Communications Letters*, vol. 6, no. 6, pp. 706–709, Dec 2017.